# Title: Polarization-multiplexed, dual-beam swept source optical coherence tomography angiography

*Jianlong Yang, Rahul Chandwani, Rui Zhao, Zhuo Wang, Yali Jia, David Huang, and Gangjun Liu\**

\*Corresponding Author: E-mail: liga@ohsu.edu

Casey Eye Institute, Oregon Health and Science University, 3375 SW Terwilliger Blvd, Portland, Oregon, 97239, USA

A polarization-multiplexed, dual-beam setup is proposed to expand the field of view for a swept source optical coherence tomography angiography (OCTA) system. This method used a Wollaston prism to split sample path light into two orthogonal-polarized beams. This allowed two beams to shine on the cornea at an angle separation of ~ 14 degrees, which led to a separation of ~ 4.2 mm on the retina. A 3-mm glass plate was inserted into one of the beam paths to set a constant path length difference between the two polarized beams so the interferogram from the two beams are coded at different frequency bands. The resulting OCTA images from the two beams were coded with a depth separation of ~ 2 mm. $5 \times 5$ mm$^2$ angiograms from the two beams were obtained simultaneously in 4 seconds. The two angiograms then were montaged to get a wider field of view (FOV) of ~ $5 \times 9.2$ mm$^2$.

## 1. Introduction

Optical coherence tomography angiography (OCTA) is an emerging imaging modality that can be used to image microvasculature noninvasively [1-2]. However, commercially available OCTA systems usually have an A-line speed of 70 ~ 100 kHz and the field of view (FOV) of high-resolution OCTA is limited. A faster system could be used to obtain wider FOV OCTA

[3]. However, the degraded signal to noise ratio (SNR) could not allow clear visualization of capillaries. Currently, the most popular approach increasing the FOV of OCTA is to montage multiple scans from different areas acquired at different times [4-6]. However, this method can be time-consuming and impractical in some clinical practices.

Two or more scan beams have been employed in many OCT applications. In the 1990s, a dual-beam setup enabled high-precision biometry of the human eye *in vivo* [7]. Dual and multiple beam setups have been used to increase the effective A-line rate of the OCT system [8-10]. Multiple beam systems have also been utilized in Doppler OCT to measure the absolute retinal blood flow rate or total retinal blood flow [11-13]. Makita *et al.* demonstrated a time-delayed dual beam system for Doppler OCTA with high sensitivity and high speed [14]. Here, we demonstrate a polarization-multiplexed dual-beam system to expand the FOV of OCTA. The system only uses one swept laser, one interferometer, and one balanced detector. Two probe beams with two orthogonal polarization states are used to image different regions of the retina simultaneously. The interferograms from the two probe beams are detected with a single detector and coded at different frequency bands. The OCTA images from the two probe beams are montaged to obtain a wider FOV OCTA image.

## 2. Materials and Methods
### 2.1. System Configuration and Characterization

Figure 1(a) shows the schematic of the dual-beam OCTA system. We employed a swept source laser (Axsun Technologies, Billerica, MA, USA) with an A-line speed of 100 kHz, a central wavelength of 1050 nm, and a tuning range of 110 nm. The theoretical axial resolution of the system is ~5.7 μm. The built-in k-clock of the swept source laser enabled an axial imaging range of ~3 mm, which was insufficient for our dual-beam OCT setup. Thus, we adopted a frequency doubling circuit [6, 15, 16] to increase the axial imaging range to ~6 mm.

A 75:25 fiber coupler (AC Photonics, Santa Clara, CA, USA) was used to split the laser light into the sample arm (25%) and reference arm (75%). In the reference arm, a fiber-coupled optical delay line was used to fine tune the optical path length difference between the two arms and an iris diaphragm (D5S, Thorlabs, Newton, NJ, USA) was mounted on the delay line to control the reference power. A 50:50 fiber coupler (TW1064R5F2A, Thorlabs, Newton, NJ, USA) was used for combining the reference light and the back-scattered sample light, which ultimately were sent into a balanced detector (PDB481C-AC, Thorlabs, Newton, NJ, USA). The interferometric signal was digitalized by a high-speed digitizer (ATS9350, AlazarTech, Pointe-Claire, QC, Canada) and then processed by a computer.

In the sample arm, the light output from the fiber was collimated by a collimator and then passed through an electronically tunable lens (EL-10-30-C, Optotune, Dietikon, Switzerland), which was used tune the focal plane location. A fast automatic search algorithm was used for rapid focusing [16]. After passing the tunable lens, the laser light traveled through a dual beam manipulation component to achieve two spatially-separated orthogonally-polarized beams. The dual beam manipulation component included a Wollaston prism and two golden mirrors (Figure 1b). The Wollaston prism (WP10, Thorlabs, Newton, NJ, USA) separated the collimated light into two orthogonally polarized outputs with a separation angle of 20°. The two 2-inch gold mirrors (PF20-03-M01, Thorlabs, Newton, NJ, USA) converged the two beams with an 6° separation angle. A 3-mm glass plate (WG10530-C, Thorlabs, Newton, NJ, USA) was inserted into one of the beams to set an optical path length difference of ~ 2 mm between two probe beams. After passing a two-axes galvo scanner (6215H, Cambridge Technology, Bedford, MA, USA) and a telescope system, the beams were able to shine on the cornea at a separation angle of 14 degrees. The telescope included a scan lens (AC508-100-B, Thorlabs, Newton, NJ, USA) with a focus length of ~100 mm and a compound ocular lens (two 49-380, Edmund Optics Inc., Barrington, New Jersey) with an effective focal length of ~40 mm. Since the two probe beams shared the same reference arm and had an optical path length difference of ~ 2 mm, the images from the two probe beams were coded in the same image at different depths. The right side of

Fig. 1(a) shows an example of the acquired retinal B-scan OCT image overlaid with its angiogram.

To guarantee the performance similarity of the two probing beams, two polarization controllers (FPC030, Thorlabs, Newton, NJ, USA) were inserted, respectively, in the sample and reference arms. In the sample arm, the polarization controller was used to even the probe powers of the two orthogonally-polarized beams. In the reference arm, the polarization controller optimized the image quality by matching the polarization states to the sample arm. The beam profiles for the two probe beams were measured by a scanning slit optical beam profiler (BP209-IR, Thorlabs, Newton, NJ, USA) (Figure 2a). The results show a full-width-at-half-maximum beam diameter of ~ 1 mm for both beams. The sensitivity roll-off performance for the two beams are shown in Figures 2b and 2c. The measured peak sensitivities for the two beams were 95.5 dB and 95.9 dB, respectively. The 6-dB roll-off depth for both beams was ~ 4.3 mm. These results show a comparable performance for the two beams.

**2.2. Scan Schematic and Laser Safety**

To apply this method for imaging the human retina with the goal of expanding the FOV of OCTA, the scanning scheme and protocol have to be carefully selected. Figure 3 shows the scan schematic and protocol of the dual-beam OCTA system. As shown in Fig. 3, the corresponding distance between the two focal spots on the retina was ~ 4.2 mm. A scanning protocol of 500×2×400 A-lines (500 A-lines per B-scan, two repetitions at the same B-scan location and 400 B-scan locations) with a FOV of 5×5 mm$^2$ was used. This separation angle (and the distance between the two focal spots on the retina) and scanning protocol was specifically chosen based on the system speed and the total scanning time (i.e. 4 seconds) for a single 3D volumatic angiography scan. Because the separation between the two beams was 4.2 mm and the total scanning size for a single beam was 5×5 mm$^2$, there was an overlap area of 0.8×5 mm$^2$ between the two areas scanned by the two beams. Therefore, the total scanning area by the two beams was 5×9.2 mm$^2$. To calculate the OCTA signal, we used two repeated B-scans at the same FOV. The total scan time was 4 seconds and the transverse sampling step size was ~12.5 μm for both directions.

A power of 1.6 mW was used for each of probe beams. The two probe beams had a separation of ~ 14 degrees (244 mrad) on the retina, which is larger than the "limiting angle of acceptance" (100 mrad) set by the International Standard for Safety of Laser Products (IEC 60825-1) [17]. Thus, the two beams were treated independently when considering ocular laser safety. The 1.6 mW power of each beam is below the most conservative limit for the 1-μm laser radiation used in retinal imaging (1.9 mW) [8]. On the cornea, the two beams were overlapped and the total power was 3.2 mW. This power is below the laser safety limit for anterior segment that was set by the American National Standard for the Safe Use of Lasers (ANSI Z136.1-2014) [18]. This concludes that the power was safe for both the anterior segment and the retina according to the IEC 60825-1 [17] and ANSI Z136.1-2014 [18]. For these reasons, the system has been approved by the Oregon Heath & Science University Institutional Review Board for clinical usage. All the studies performed using this system follow the tenets of the Declaration of Helsinki for the treatment of human subjects.

## 3. Results

We demonstrate the results of this dual-beam OCTA system in Fig. 4 and Fig. 5. A healthy male volunteer of age 29 years was involved in this research. Retinal fovea and optic nerve head (ONH) regions was imaged. Figure 4 shows the 3D intensity volumes (Figures 4a and 4d) and cross-sectional images (Figures 4b, 4c, 4e, 4f). In the cross-sectional images, red-color angiograms were overplayed on gray-scale reflectance images. It can be seen that the two beams delivered very similar performance. Since the scan regions of the two beams have an overlapping of ~0.8×5 mm, they were able to be effectively montaged. Figure 5 shows the montaged *en face* angiograms of fovea (a) and ONH (b). The red dash lines in the middle indicate the stitching location of the images.

The current setup had a separation of ~ 4.2 mm on human retina, which can be altered by properly tuning the distance between the Wollaston prism and the two gold mirrors as well as tilting the angles of the two mirrors. This could allow more versatile setups for other applications such as the time-delayed dual beam Doppler OCT and OCTA [14]. In addition, the current probe beams were separated vertically on the retina and they simultaneously image the retina superiorly and inferiorly. By properly rearranging the Wollaston prism and two mirrors, the two beams can be separated horizontally on the retina. This could allow for imaging the macular and optic disk regions simultaneously in a single scan.

The polarization states of the reference arm cannot be perfectly aligned with those of the sample arm by manually tuning the PC, which may degrade the detection sensitivity [9]. Because non- polarization maintained fiber was used in this setup, the polarization state was not always maintained. This requires re-tuning of the polarization controllers for long term usage of the system. We are developing an automatic tuning method based on electrically tunable PC to improve the performance.

## 4. Conclusion

In summary, we demonstrated a dual-beam swept-source OCT system based on polarization-multiplexing and employed this method to enlarge the FOV of OCTA. A Wollaston prism was used to create two orthogonally-polarized beams while a thin glass sheet allowed for separating the images from these two beams in a single B frame. Our system avoided the use of multiple interferometers and detectors, plate and cube polarizing beam splitters, and related extra optics, which substantially simplified the system's design. The two beams showed very similar performance and gave comparable OCT image quality. By montaging two OCTA volumes that were simultaneously captured in 4 seconds, we obtained *en face* angiograms of the human retina with a FOV of $5 \times 9.2$ mm$^2$. These results show the feasibility of further applying this design in faster OCT systems to shorten the acquiring time of wide-field OCTA.

**Financial Disclosure**    This research was funded by Oregon Health & Science University Foundation, National Institutes of Health Grants DP3 DK104397, R01 EY024544, R01 EY023285 and R01 EY018184, unrestricted departmental funding from Research to Prevent Blindness (New York, NY), and P30 EY010572 from the National Institutes of Health (Bethesda, MD).

**Funding**    This research was funded by Oregon Health & Science University Foundation, National Institutes of Health Grants DP3 DK104397, R01 EY024544, R01 EY023285 and R01 EY018184, unrestricted departmental funding from Research to Prevent Blindness (New York, NY), and P30 EY010572 from the National Institutes of Health (Bethesda, MD).

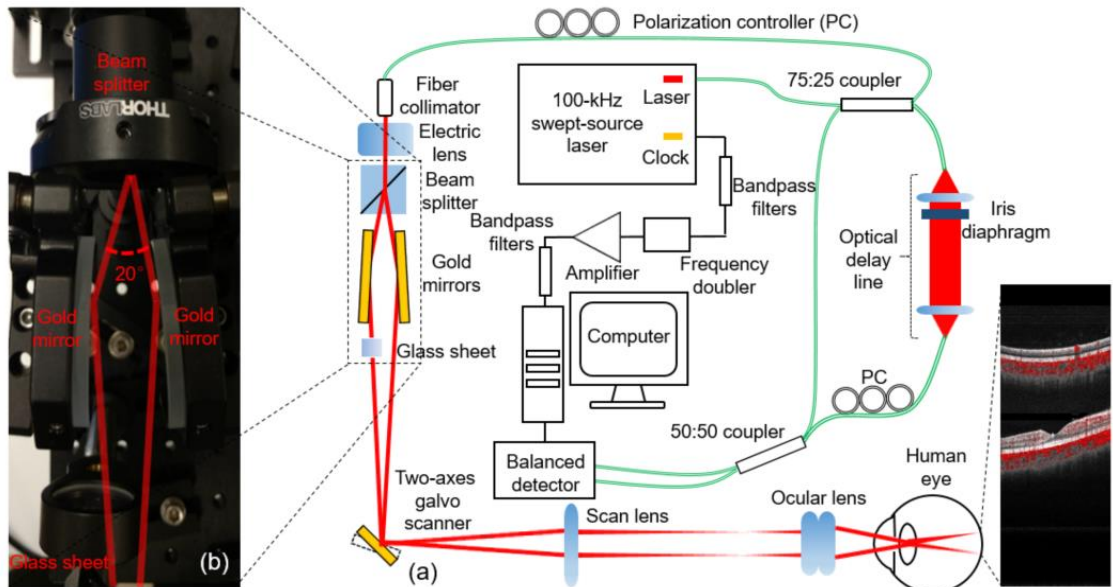

**Figure 1.** (a) Schematic of the dual-beam OCTA system. (b) Photograph of the dual beam manipulation components.

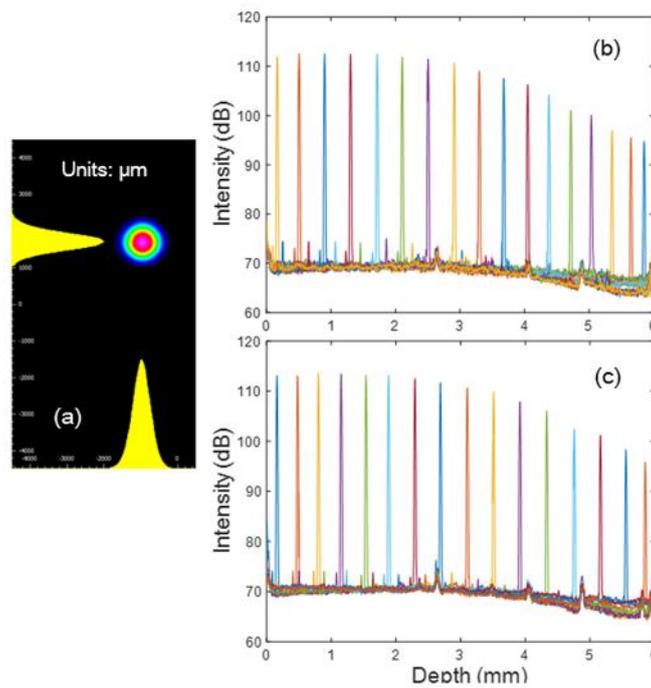

**Figure 2.** (a) 2-dimensional beam profiles at the position of cornea. (b) and (c) are the sensitivity roll-offs of the two orthogonally-polarized probe beams.

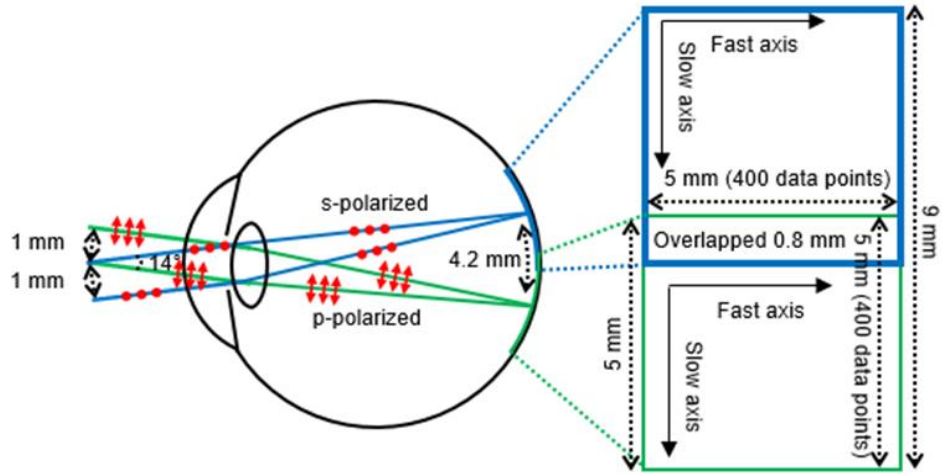

**Figure 3.** Scan schematic and protocol of the dual-beam OCTA system.

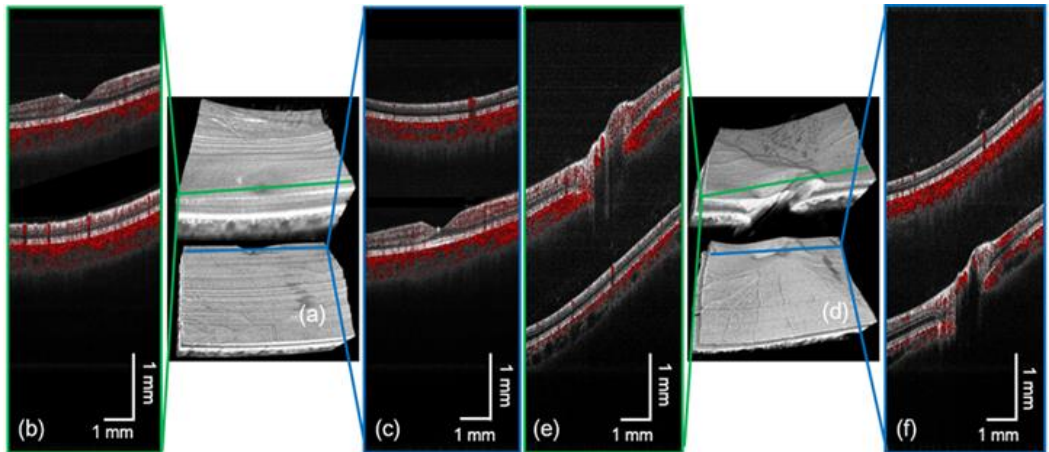

**Figure 4.** Captured OCTA image data sets at around fovea and optic nerve head (ONH). (a) and (d) show the 3D intensity volumes. (b), (c), (e), and (f) show the cross-sectional intensity images (gray) overlaid with angiograms.

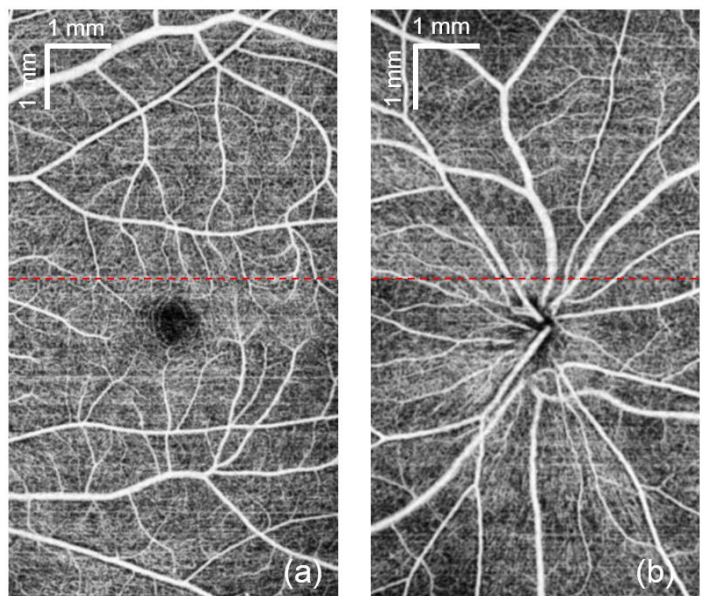

**Figure 5.** Montaged en face angiograms of retina regions at around (a) fovea and (b) optic nerve head. The red dash lines in the middle give the positions of stitching images.

# Graphical Abstract

We demonstrated a dual-beam swept-source OCT system based on polarization-multiplexing to enlarge the FOV of OCTA. A Wollaston prism was used to create two orthogonally-polarized beams. By montaging two OCTA volumes that were simultaneously captured in 4 seconds, we obtained *en face* angiograms of the human retina with a FOV of $5 \times 9.2$ mm$^2$.

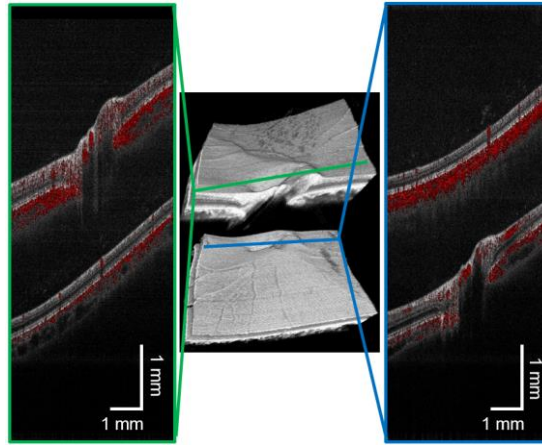